\documentclass[11pt]{article}
\usepackage{latexsym}
\usepackage[mathscr]{eucal}
\usepackage{amscd}
\usepackage{amssymb}
\usepackage{amsmath}

\renewcommand{\u}{{\bf u}}
\renewcommand{\v}{{\bf v}}

\renewcommand\S{\Sigma}
\newcommand\s{\sigma}

\newcommand\e{\epsilon}
\renewcommand\b{\beta}

\newcommand\g{\gamma}
\newcommand\8{\infty}
\renewcommand\a{\alpha}

\newcommand\beq{\begin{eqnarray}}
\newcommand\eeq{\end{eqnarray}}
\newcommand\ben{\begin{enumerate}}
\newcommand\een{\end{enumerate}}
\newcommand\beqq{\begin{eqnarray*}}
\newcommand\eeqq{\end{eqnarray*}}

\usepackage{verbatim}
\newcommand{\tR}{\widetilde R}
\newcommand{\tSi}{\widetilde \Sigma}
\newcommand{\tSigma}{\widetilde \Sigma}

\newcommand{\tN}{\widetilde N}

\newcommand{\tn}{\widetilde n}
\newcommand{\wh}{\widetilde h}

\newcommand{\tD}{\widetilde \Delta}
\newcommand{\tW}{\widetilde W}

\newcommand{\tS}{\widetilde S}

\newcommand{\tb}{\widetilde b}

\newcommand{\tnabla}{\widetilde \nabla}
\newcommand{\tH}{\widetilde H}

\newcommand{\qed}{\hfill $\Box$ \medskip}

\newcommand{\re}{{\mathbb R}}

\newcommand{\pd}{\partial}

\newcommand{\cR}{\widetilde {\cal R}}
\newcommand{\cA}{{\cal A}}

\newcommand{\cH}{{\cal H}}

\newcommand{\cU}{{\cal U}}

\newcommand{\cO}{{\cal O}}

\newcommand{\cV}{{\mathcal V}}

\newcommand{\cK}{{\mathcal K}}

\newcommand{\fg}{{\mathfrak g}}

\newcommand{\cT}{{\mathcal T}}
\newcommand{\sinf}{\Sigma_\infty}
\newcommand{\sh}{\Sigma_H}
\newcommand{\tP}{{\widetilde P}}
\newcommand{\tQ}{{\widetilde Q}}

\newcommand{\mD}{{\mathcal D}}

\makeatletter
\renewcommand{\section}{\@startsection{section}{1}{0mm}%
{-0.5\baselineskip}{1 pt}{\normalfont\large\bfseries}}
\renewcommand{\subsection}{\@startsection{subsection}{2}{0mm}%
{-0.1\baselineskip}{1 pt}{\normalfont\bfseries}}
\renewcommand{\subsubsection}{\@startsection{subsubsection}{3}{0mm}%
{-0.1\baselineskip}{1 pt}{\normalfont\bfseries}}
\renewcommand{\paragraph}{\@startsection{paragraph}{4}{0mm}%
{-0.1\baselineskip}{1 pt}{\normalfont\itshape}}
\renewcommand{\subparagraph}{\@startsection{subparagraph}{5}{0mm}%
{-0.1\baselineskip}{1 pt}{\normalfont}}
\makeatother

\title{Non-Existence of Black Holes in \\ Certain  $\Lambda<0$
Spacetimes.}

\author{G.J.\ Galloway${}^{a}$\footnote{galloway@math.miami.edu}, S.\
Surya${}^{b,c}$\footnote{e-mail: ssurya@phys.ualberta.ca} , E.\
Woolgar${}^{c,b}$  \footnote{e-mail: ewoolgar@math.ualberta.ca}
\\
\\
${}^a$Dept.\ of Mathematics \\
University of Miami, Coral Gables, FL 33124, USA
\\
\\
${}^b$Theoretical Physics Institute, University of Alberta\\ Edmonton,
AB, Canada T6G 2G1
\\
\\
${}^{c}$Dept.\ of Mathematical Sciences, University of Alberta\\
Edmonton, AB, Canada T6G 2G1}
\date{\today}

\begin{document}
\maketitle
\begin{abstract}
Assuming certain asymptotic conditions, we prove a general theorem on
the non-existence of static regular black holes in spacetimes with a
negative cosmological constant, given that the fundamental group of
space is infinite.  We use this to rule out the existence of regular
negative mass AdS black holes with Ricci flat scri.  For any mass, we
also rule out a class of conformally compactifiable static black holes
whose conformal infinity has positive scalar curvature and infinite
fundamental group, subject to our asymptotic conditions. 
In a limited, but important, special case our result adds new 
support to the AdS/CFT inspired  positive mass conjecture of Horowitz and
Myers.  
\end{abstract}
\vspace{0.1 cm}

\section{Introduction}

The discovery of new classes of asymptotically Anti-de Sitter
(AdS) black holes in the past decade
\cite{lemos,brill,swedish,vanzo,db,mann} has brought to light
the fact that spacetimes with negative cosmological constant can admit
black hole event horizons of nonspherical spatial topology. This is
possible because these black holes violate neither of the two known
obstructions to nontrivial horizon topology. The first of these is the
theorem of Hawking \cite{SWH1}, which assumes an energy condition
incompatible with a negative cosmological constant and so does not
apply. The second is topological censorship \cite{FSW,GSWW}, which
applies regardless of the cosmological constant but is not in itself a
constraint on horizon topology. It merely enforces a topological
relation between the horizon and the Penrose conformal boundary scri
(when there is one), which consequently is also spatially nonspherical
for black holes with nonspherical horizons.
\footnote{By ``nonspherical scri'' we mean ``spatially
nonspherical''; {\it i.e.}, slices of scri orthogonal to its 
timelike conformal Killing field are not topological spheres.}

Despite differing from anti-de Sitter spacetime in the global
structure of scri, these spacetimes still satisfy fall-off
conditions that are essentially those given by Henneaux and
Teitelboim \cite{hennteit, MH} (see also \cite{SWH2,am,ashdas})
for asymptotically anti-de Sitter spacetimes, and so serve as
examples of what are called asymptotically {\it locally} AdS
(AL-AdS) spacetimes. Spacelike slices of scri are homogeneous
manifolds of either positive, zero, or negative Ricci curvature.
Interestingly, when these have negative Ricci curvature, the
associated black holes can have negative mass \cite{db,mann}.

In this paper, we concern ourselves with related black hole existence
issues. These are motivated by a curious property of asymptotically
locally AdS spacetimes with nonspherical scri, namely that the lowest
energy solutions or ground states need not be locally AdS ({\it i.e.},
constant curvature).  An example of a nonsingular, nonconstant
curvature spacetime with Ricci flat conformal boundary which has {\it
negative mass} (relative to a zero of mass corresponding to constant
curvature) is the AdS soliton \cite{witten98,hm}. Indeed, assuming a
generalisation of the AdS/CFT correspondence conjecture \cite{adscft,
gkp,witten98} which states that string theory on the $5$ dimensional
AdS soliton is equivalent to a certain non-supersymmetric gauge theory
on scri, Horowitz and Myers conjectured that, in $5$ spacetime
dimensions, the AdS soliton is the ground state in its asymptotic
class \cite{hm}.  In \cite{gsw1,gsw2} we proved (in all spacetime
dimensions $\ge 4$) that the AdS soliton is the unique negative mass
{\it globally} static vacuum solution satisfying certain asymptotic
and topological conditions.  In other words, there is a ``mass gap''
in this class of static solutions. Dropping the requirement of global
staticity allows us to include black hole regions in the spacetime (we
retain the assumption of an irrotational timelike Killing field in the
domain of outer communications: such spacetimes are called static but
not globally static).  However, as we will show, there exist no
regular, static, negative mass black holes with Ricci flat scri, so
that our uniqueness result holds more generally. This result is
summarised in:

\noindent {\bf Proposition 1.1} {\sl Let $(\Sigma, h, N)$ be an $n+1$
dimensional ($n\geq 3$) static AL-AdS spacetime which satisfies the
null energy condition
$R_{ab}k^ak^b\ge 0$ $\forall$ null $k^a$,
has Ricci flat scri, and has Ashtekar-Magnon mass aspect pointwise
negative on scri. If $\pi_1(\Sigma)$ is infinite and if the
asymptotic condition (S) (Definition 2.1) is satisfied, then
$(\Sigma, h, N)$ does not admit regular black holes.}

Here (and in (1.2) and (1.3)), $(\Sigma,h,N)$ refers to a spacetime 
with static domain of outer communications of the form 
$(\re \times \Sigma, -N^2 dt^2 \oplus h)$. By ``regular black hole'', 
we mean one with a smooth, nondegenerate event horizon of any topology, 
cf., Section 2 for further details.

The uniqueness theorem of \cite{gsw1,gsw2} can be interpreted as
evidence for the Horowitz-Myers conjecture for $n=4$. The above
Proposition strengthens this evidence, by extending the uniqueness to
the class of non-nakedly-singular spacetimes with an irrotational
Killing field that is timelike near infinity (together with certain
asymptotic conditions).  This class of spacetimes is important, since
if there is a ground state, it should be static (though not
necessarily globally static), i.e. should have zero kinetic energy.

Anderson, Chru\'sciel, and Delay \cite{ACD} have recently
established the uniqueness of the AdS soliton and the Lemos
toroidal black holes \cite{lemos} in the $n=3$ ({\it i.e.},
spacetime dimension $4$) case with fewer asymptotic restrictions. It would
be interesting from the point of view of physics to extend their
degree of generality and their methods to $n=4$.

When the conformal boundary of $\Sigma$ at infinity, $\pd \sinf$, has
generic topology and geometry, the existence of solutions (let alone a
well defined ground state) is largely an open question (cf., however,
\cite{ACD}). If a regular static solution were to exist for a given
$\pd \sinf$, one might expect it to also admit black holes.  The
following proposition establishes further restrictions on the models
that can occur as static AL-AdS spacetimes.

\noindent {\bf Proposition 1.2.} {\sl Let $(\Sigma, h, N)$ be an
$(n+1)$-dimensional ($n\geq 3$ ) static AL-AdS spacetimes
satisfying the null energy condition. Suppose (a) the conformal
boundary at infinity $\pd \sinf$ has positive scalar curvature
(with respect to the {\it Fermat metric}, see Section 2), (b) the
fundamental group of $\S$ is infinite,
$\vert \pi_1(\S)\vert = \8$,

and (c) the asymptotic condition (S) (Definition 2.1) is
satisfied. Then $(\Sigma, h, N)$ does not admit regular black holes.}

What sort of conformal boundary $\pd\sinf$ is compatible with the
assumptions (a--c)? In $4+1$ dimensions the prototypical example
is $\pd\sinf =  S^1\times S^2$, with standard metric. In general,
topological censorship \cite{GSWW} requires that $\pd \sinf$ have infinite
fundamental group whenever $\S$ does.  Moreover, as follows from
the boundary analysis in Section 3 (see also Appendix A),
assumptions (a) and (c) imply that $\pd \sinf$ must have
nonnegative Ricci curvature. (Further connections between the
convexity condition (S) and the curvature of the conformal
boundary are considered in Section 3). Thus Proposition 1.2 rules
out, within a certain class of static AL-ADS spacetimes, conformal
boundaries which, for example, are products of flat spaces and
spaces of positive Ricci curvature.

One may ask whether one can use the above results to rule out certain
horizon topologies (once again, our term ``black hole'' includes
nonspherical horizon topologies).  This is possible if one restricts
to the class of ``product'' black hole spacetimes, namely ones whose
domain of outer communications has the simple product topology $\re
\times \pd \sinf \times \re^+$.  For such spacetimes, scri and the
horizon are diffeomorphic. Proposition 1.1 then implies that, under
certain asymptotic conditions, there are no (pointwise) negative mass,
regular, static, product black hole spacetimes with horizon topology
$\re^{n-1}/\Gamma$, with $\Gamma$ a discrete co-compact isometry group
of $\re^{n-1}$. In particular, it rules out static (pointwise)
negative mass toroidal horizons in product black hole spacetimes
(consistent with \cite{ACD} in the $n=3$ case, wherein it is shown
that the positive mass toroidal black holes of Lemos \cite{lemos} are
unique amongst vacuum solutions).  Similarly, as follows from the
discussion above, Proposition 1.2 rules out, for example, a certain
family of product spacetimes with horizon topology $U\times V$ where
$U=S^{k_1} \times \ldots S^{k_m}$ for $k_i \geq 2, \, \, \, m \geq 1$
and $V=S^1 \times
\ldots S^1$. In $n+1=5$ it thus rules out static product black
hole spacetimes with $S^2 \times S^1$ horizons.

Our proof makes crucial use of the construction of a complete
achronal null geodesic, or {\sl null line} in \cite{gsw1,gsw2} and
the subsequent use of an extension of the null splitting theorem
(NST) of Galloway \cite{gjg} (Theorem 3.1). The NST was used to show 
that the existence of a null line implies a certain rigidity of the
spacetime---indeed, under certain asymptotic restrictions the only
vacuum spacetimes with zero and positive cosmological constants
admitting null lines
are Minkowski and de Sitter spacetimes, respectively
\cite{gjg,gjg2}. On the other hand, 
as we will show, null lines always exist in the universal cover of a
static AL-AdS spacetime whenever this cover contains non-compact spatial
directions, given our asymptotic conditions. The presence of regular
black holes regions in the spacetime, however, implies the existence
of strictly convex surfaces (in the ambient Fermat metric) in the
neighbourhood of the black hole, which can be shown, using the NST, to
be incompatible with the presence of a null line.  Thus, as we will
prove, the existence of a null line in the interior of a static
spacetime is incompatible with the existence of regular black holes.

In section \ref{preliminaries}, we begin by specifying the class of
spacetimes which we will consider.  Apart from the requirement that
the spacetime be AL-AdS
and static (in the domain of outer communications),
we further require it to satisfy a certain
asymptotic convexity condition, which we denote as Condition (C).
Condition (C) can be related to the mass when the boundary is Ricci
flat, and shown to be equivalent to a weaker condition (S) when the
boundary has positive Ricci curvature. This involves a boundary
analysis along the lines of \cite{FG} which we include in the
Appendix.
Next, we observe that there exists a neighbourhood of any
regular horizon, in which all constant lapse surfaces outside the
horizon are strictly convex with respect to the Fermat metric.
In section \ref{mainresults} we
first show that the null generators of the boundary
of the past (respectively future) of the null line
are future (respectively past) complete. 
This fact then allows us to use the null splitting theorem 
to prove our black hole non-existence theorem:

\noindent {\bf Theorem 1.3.} {\sl Let $(\Sigma, h,N)$ be an $n+1$
dimensional ($n\geq 3$) static AL-AdS spacetime satisfying the
null energy condition. If $\pi_1(\Sigma)$ is infinite and if the
asymptotic condition (C) (Definition 2.1) is satisfied, then
$(\Sigma, h, N)$ does not admit regular black holes.}

\noindent
In Section 4, we show that under the stated assumptions, 
Propositions 1.1 and 1.2 follow as corollaries of Theorem 1.3.

\section{Preliminaries}
\label{preliminaries}
\setcounter{equation}{0}

We will consider $(n+1)$-dimensional, $n\ge 2$, {\sl AL-AdS
spacetimes} $M$,
with an event horizon ${\cal H}$ (possibly empty). Let $(\mD,g)$
be a static connected component of the domain of outer
communications, so there is an irrotational timelike Killing
vector field on $(\mD,g)$, which extends smoothly to 
$\partial \mD={\overline \mD} \cap{\cal H}$, where it becomes null. 
Then we have
\begin{equation}
\mD = \re \times \Sigma\quad ,\quad  g= -N^2dt^2 \oplus h\quad ,
\quad N|_{\partial \mD}=0\quad , \label{FE1}
\end{equation}
where $h$ is the induced metric on $\Sigma$ and $N$ is the lapse,
such that the triple $(\Sigma, h, N)$ is conformally
compactifiable. For technical reasons, we will assume that the
static black hole horizon is {\it regular} (nondegenerate), {\it
i.e.}, $d N|_{\cH}\neq 0$ pointwise. Thus, there exists a smooth
compact manifold with boundary, $\Sigma'=\Sigma \cup \pd \sh \cup
\pd \sinf$, where $\pd \sh= \Sigma' \cap \cH$, such that \ben

\vspace{-.1in}
\item[(a)] $N$ and $h$ extend smoothly to the
closure $\overline \Sigma =\Sigma \cup \pd \sh$ of $\Sigma$ in
$M$,  with  $N|_{\pd
\sh}=0$, $dN|_{\pd \sh} \neq 0$,
\item[(b)] $N^{-1}$ extends to a smooth
function $\tN$ on $\tSi \equiv\Sigma'\backslash \pd \sh$, with
$\tN|_{\pd \sinf}=0$ and $d\tN|_{\pd \sinf} \neq 0$ pointwise, and
\item[(c)]
$N^{-2}h$ extends to a smooth Riemannian metric $\wh$, the {\it Fermat
metric} on $\tSi$.
\een

\vspace{-.1in} We will refer to the triple $(\Sigma, h, N)$ as the
{\sl physical spacetime}, and $(\tSi, \wh, \tN)$ as the {\sl Fermat
conformal gauge}.  Note in this gauge, when $\pd \sh \neq \emptyset$,
the surface $N=0$ gets mapped to infinity.  We allow $\pd \sh$ to have
multiple components.

We also require  $(\Sigma, h, N)$ to satisfy the static field
equations,
\begin{eqnarray}
R_{ab}& = &N^{-1}\nabla_a\nabla_b N +\frac{2\Lambda}{n-1}
h_{ab}+\cT_{cd}h^c_ah^d_b
\label{FE2}, \\
\Delta N & = & -\frac{2\Lambda}{n-1} N + \frac{1}{N}\cT_{00},
\label{FE3}
\end{eqnarray}
where $\nabla_a$ and $R_{ab}$ are respectively the covariant
derivative and Ricci tensor on $(\Sigma, h)$, the Laplacian is
$\Delta=h^{ab}\nabla_a\nabla_b$, the cosmological constant is
$\Lambda <0$, and
\begin{equation}
\cT_{ab}=T_{ab}- \frac{1}{n-1}g_{ab}g^{cd}T_{cd},
\end{equation}
with $T_{ab}$ the matter stress-energy tensor, such that $N^{n-1}
T_{ab}$ admits a smooth limit to $\pd \sinf$.

It will be useful for
the purposes of Section 3 and the appendix to write the field
equations in terms of the Fermat metric $\wh$ and associated
$\tnabla_a$ and $\tR_{ab}$,
\begin{eqnarray}
\tR_{ab}&=&-\frac{(n-1)}{\tN}\tnabla_a\tnabla_b \tN + 8 \pi \cT_{cd}
\wh^c_a \wh^d_b\quad ,\label{A1}\\
\tN \tD\tN&=&\left (\frac{2\Lambda}{n-1}+n\tW\right ) -8 \pi \tN^2
\cT_{00}\quad ,\label{A2}
\end{eqnarray}
where
\begin{equation}
\tW:={\wh^{ab}}\tnabla_a\tN\tnabla_b\tN =\frac{1}{N^2}h^{ab}\nabla_a
N\nabla_b N\label{A3}.
\end{equation}

For our theorem we will need to examine the behaviour of geodesics
in the neighbourhood of the compact boundary $\pd \Sigma'=\pd \sh
\cup \pd \sinf$. Our analysis will make use of  the existence of a
``collar neighbourhood'' of the boundary $U=\pd \Sigma' \times [0,
\epsilon)$ whose interior is foliated by constant lapse surfaces.
Such a collaring exists for any function $f$ satisfying (i)
$f|_{\pd \Sigma'}= const$ and (ii) $ df|_{\pd \Sigma'} \neq 0$.
Putting $f=N$ for the regular black hole boundary $\pd \sh$, and
$f=\tN$ for the boundary at infinity $\pd \sinf$ and using
conditions (a), (b) and (c) above, we may therefore construct
collar neighbourhoods of both $\pd \sh$ and $\pd \sinf$ whose
interiors are foliated by constant lapse surfaces $\cV_c=\{N =
c\}$.

We will require a particular convexity assumption for the constant
lapse surfaces in a neighbourhood of scri in the Fermat conformal
gauge (see also \cite{gsw2}). Let us consider
a collar  neighbourhood of $\pd \sinf$ which is foliated by the
level surfaces ${\cV_c} = \{N = c\}$ of the lapse $N$. The second
fundamental form of the $\cV_c$ in the Fermat conformal gauge,
with respect to the unit normal vector $\tn^a$ pointing towards
scri, is given by $\tH_{ab}=(\tnabla_{a} \tn_{b})_{\perp}$ where
$\perp$ denotes the projection to the $\cV_c$. The eigenvalues of
$\tH_{ab}$ are called the {\it principal curvatures} of the
$\cV_c$.

\noindent {\bf Definition 2.1.} We say that $(\Sigma,h,N)$
\emph{satisfies condition (S)} provided that the second
fundamental form $\tH_{a b}$ of each level surface $N=c$ is
semi-definite (equivalently, provided that the principal
curvatures of each level surface $N=c$ are either all non-negative
or all non-positive) whenever $c$ is sufficiently large ({\it
i.e.}, near scri). If $\tH_{a b}$ is positive semi-definite
(equivalently, if the principal curvatures are all non-negative)
for each of the level surfaces in this neighbourhood of scri, we
say that $(\Sigma,h,N)$ \emph{satisfies condition (C)}, and the
level surfaces of $N$ in this neighbourhood are said to be {\it
weakly convex}.

As in \cite{gsw2}, Condition (C) will be used to control the
behaviour of certain geodesics near scri as follows.  Suppose
Condition (C) holds, so that the level surfaces $\cV_c:=\{N=c\}$
are weakly convex, in the sense of the definition, for all $c$
sufficiently large.  Let $\cV_0= \{N=c_0\}$ be such a level
surface; $\cV_0$ has a well-defined ``inside'' ($N<c_0$) and
``outside'' ($N>c_0$). Then, as follows from the maximum
principle, if $\gamma$ is a geodesic segment with endpoints inside
$\cV_0$, all of $\gamma$ must be contained inside $\cV_0$.  Thus,
Condition~(C) provides ``barrier surfaces'' for the construction
of certain minimizing geodesics, as will be seen in the next
subsection. Condition (S) on the other hand allows the level
surfaces $N = c$ near scri to be either weakly convex ($H_{a b}$
positive semi-definite) or weakly concave ($H_{a b}$ negative
semi-definite).  All the relevant examples known to us obey
condition (S), even when Condition (C) fails. As we will see in
Section 4, Condition (S) along with certain extra boundary
conditions may be used to obtain Condition (C).

We also need to understand the behaviour of geodesics near the
black hole boundary $\pd \sh$. In the following we will call a
surface $\cK \subset \Sigma$ {\it strictly convex} with respect to
a choice of normal if the second fundamental form $\tH_{ab}$ in
the Fermat conformal gauge is positive definite (equivalently, its
principal curvatures are strictly positive) with respect to this
normal.

\noindent {\bf Lemma 2.2.} {\sl Let $(\Sigma, h, N)$ be a static
spacetime with regular black hole boundary $\pd\sh \neq
\emptyset$.  Then for each $c>0$ sufficiently small, the constant
lapse surface $\cV_c = \{N=c\}$ is diffeomorphic to $\pd \sh$ and
strictly convex in $(\tSi, \wh)$ with respect to the normal
pointing towards  $\pd \sh$.  Thus, any Fermat geodesic with
endpoints in the region $N>c$ cannot intersect $\cV_c$.}

 \noindent {\bf Proof.} 
Let us consider the collar neighbourhood $\cU\approx[0,c_0)\times \pd \sh$
of $\pd \sh$ foliated
by the constant lapse surfaces $\cV_c\ =\{N=c\} \approx \pd \sh$,
$0<c<c_0$. The
second fundamental forms of each $\cV_c$ in the physical metric and in the
Fermat metric are related by
\begin{equation}
\label{KH1}
\tH_{ab}(c) = c^{-1} H_{ab}(c) - c^{-2}h_{ab}n^d \pd_d N,
\end{equation}
where $n^d$ is the unit normal field to the $\cV_c$ in the
physical metric, chosen to be pointing towards surfaces of
decreasing lapse, {\it i.e.}, towards $\pd \sh$.

Let $X^a$ be a unit vector field in the physical metric, defined
in a neighbourhood $\cO\subset \cU$ of any point $p\in\pd\sh$,
such that $X^a$ is tangent to each $\cV_c$ meeting $\cO$.  On
$\cO\setminus \pd\sh$, set $\tilde X^a = N X^a$; $\tilde X^a$ is a
unit vector field in the Fermat metric also tangent to each
$\cV_c$.  Contracting equation (\ref{KH1}) with $\tilde X^a$, we
obtain \beq \tH_{ab}(c)\tilde X^a\tilde X^b = cH_{ab}(c)X^aX^b -
n^d \pd_d N \, . \eeq Thus, in the limit as $c\to 0$,
$\tH_{ab}(c)\tilde X^a\tilde X^b \to dN(n)|_{\pd\sh}>0$. Using the
compactness of $\pd \sh$, it follows that $\tH_{ab}(c)$ is
positive definite for all $c$ sufficiently small.  The statement
regarding Fermat geodesics follows from the discussion after
Definition~2.1. \qed

\noindent {\bf Remark.} In the above proof, the condition of
regularity, $n^d\pd_d N(0) \neq 0$, is crucial in obtaining a
neighbourhood of the horizon in which $\tH_{ab}(c)$ is strictly
positive definite. For degenerate horizons, $n^d\pd_d N(0)= 0$,
and hence our lemma does not extend in a trivial manner to this
case. However, 
for familiar examples such as the extremal Reissner Nordstrom and 
the extremal charged AdS black holes, one can explicitly construct 
a strictly convex surface $\cK$ in the neighbourhood of the horizon.

\section{Black hole non-existence theorem.}
\label{mainresults}
\setcounter{equation}{0}

We begin by reminding the reader of the definition of a line.  In a
Riemannian manifold, a \emph{line} is an inextendible geodesic, each
segment of which has minimal length, while in a spacetime, a
\emph{timelike line} is an inextendible timelike geodesic, each
segment of which has maximal length.
Motivated by these more standard cases, a {\it null line} in
spacetime is defined to be an inextendible null geodesic which is
globally achronal, {\it i.e.}, for which no two points can be joined 
by a timelike curve. (Hence, each segment of a null line is  maximal
with respect to the Lorentzian arc length functional.) In static
spacetimes, spacelike Fermat geodesics in $(\tSi, \wh)$ can be
lifted, via the timelike Killing field, to null geodesics in the
physical spacetime $(\Sigma, h,N)$ under suitable
reparameterisations (see \cite{gsw2}). Indeed, there is
essentially a one-to-one correspondence between affinely
parameterised Fermat geodesics in $(\tSi, \wh)$ and affinely
parameterised, future directed null geodesics in $(\Sigma, h, N)$.
Moreover, a Fermat line in $(\tSi, \wh)$ lifts to a unique future
directed null line in $(\Sigma, h,N)$ through a fixed
basepoint~\cite{gsw2}.

Since our proof will make crucial use of a slight generalisation
of the NST, we state it here:

\noindent {\bf Theorem 3.1 (Null Splitting Theorem, Galloway
\cite{gjg}).} {\sl If a null geodesically
complete spacetime obeys $R_{ab}X^aX^b\ge 0$ for all null vectors
$X^a$ and also contains a null line $\eta$, then $\eta$ lies in a
smooth,  edgeless, totally geodesic null hypersurface.}

For our purposes, null geodesic completeness is too strong a
requirement, since we would like to allow for the presence of
black holes in the spacetime.  As was pointed out in \cite{gjg},
the NST is still valid if null geodesic completeness is dropped
and instead one imposes the less stringent requirement that the
null generators of $S^+=\pd J^+(\eta)$ and $S^-=\pd J^-(\eta)$ be
past and future complete, respectively. Indeed, it is only in
order to show this latter property of the null generators of
$S^\pm$ that null geodesic completeness was used in the proof.
Thus, in order to use the NST in the absence of null geodesic
completeness, one requires an alternative method to prove this.
For our purposes it suffices to show:

\noindent {\bf Lemma 3.2.} \label{completeness.lemma} {\sl Let
$(\Sigma,h,N)$ be a static spacetime, and let $\S_0$ be an open
subset of $\S$.  Suppose that \ben

\vspace{-.1in}
\item[(a)]  $\overline\S_0$, the closure of $\S_0$ in $\S$, is a complete
Riemannian manifold-with-boundary in the Fermat metric $\wh=
N^{-2}h$,

\vspace{-.1in}
\item[(b)] the boundary $\pd\overline \S_0$ of $\overline\S_0$ is weakly
convex (with respect to  the outward normal) in $ (\S,\wh)$, and

\vspace{-.05in}
\item[(c)] $N$ has a positive lower and upper bound on  $\S_0$.
\een \vspace{-.1in}
If $\eta$ is a complete null line in the ``truncated spacetime"
$M_0 = (\S_0, h, N)$, then the null generators of $S^+=\pd
J^+(\eta, M_0)$ are past complete, and the null generators of
$S^-=\pd J^-(\eta, M_0)$ are future complete.}

In the proof of Theorem 1.3, Lemma 3.2 will be applied to the {\it
universal cover} of the domain of outer communications.

{\bf Proof.} In what follows, pasts and futures will refer to the
spacetime $M_0 = (\S_0, h, N)$, and we identify $\S_0$ with the
time slice $t=0$ in $M_0$.

We begin with the hypersurface $S^+$ and a point $q \in S^+$.
Without loss of generality, we may assume that $q$ is not a future
endpoint of a null generator of $S^+$. We express $q=(t_0, p) \in
(\S_0, h, N)$, where $p \in \S_0$. Let $p_0\in \eta$ be the point
where $\eta$ meets $\S_0$. Let $\g$ be the projection of $\eta$
into $\S_0$; $\g$ is a complete line in $(\S_0,\wh)$ passing
through $p_0$. Let $\{q_i\}$ be a sequence of points which exhaust
$\eta$ to the past. The points $q_i=(t_i, p_i)$ with $p_i \in
\S_0$ then project to the sequence $\{p_i\}\in \gamma$ such that
the Fermat distance from $p$ to $p_i$ in $(\S_0, \wh)$ for
successive $i$ tends to infinity.  The convexity assumption
implies that any two points in $\S_0$ can be joined by a minimal
Fermat geodesic contained in $\S_0$. Thus, for each $i$, there
exists a minimal Fermat geodesic  $\sigma_i\subset \S_0$ from $p$
to $p_i$. By passing to a subsequence if necessary, the $\sigma_i$
converge to a ray (geodesic half-line) $\sigma$ in $(\S_0,\wh)$
starting at $p$; $\sigma$ is referred to as an asymptote of
$\gamma$.

Each $\sigma_i$ lifts to a unique null geodesic $\mu_i$ in the
physical spacetime with $q_i=(t_i, p_i)$ as its past endpoint. The
future endpoint of each $\mu_i$ is some $x_i=(\tau_i, p)$. We now
show that the $\tau_i$ lie in a finite interval. In $(\S_0, \wh)$,
the triangle inequality allows us to write
\begin{equation}
\label{triangle} |\tilde d(p_0,p_n) - \tilde d(p,p_n)| \leq \tilde
d(p,p_0) = l_0,
\end{equation}
where $\tilde d(p_1,p_2)$ denotes the Fermat distance between the
points $p_1$ and $p_2$. Let $\gamma_i \subset \gamma$ be a segment
of the Fermat line $\gamma$ with endpoints $p_0$ and $p_i$. Since
$\gamma_i$ and $\sigma_i$ are minimal with respect to the Fermat
metric, $\tilde d(p_0,p_i)= \tilde L(\gamma_i)$ and $\tilde
d(p,p_i)= \tilde L(\sigma_i)$, where $\tilde L(x)$ represents the
Fermat arc length.  In a static spacetime, the time taken to
traverse a null geodesic in the physical metric is given by the
arc length of the projected Fermat metric. Thus, $\tilde
L(\gamma_i)= t_i$ and $\tilde L(\sigma_i)= t_i -\tau_i$, which,
from (\ref{triangle}) implies that $|\tau_i| \leq l_0$. Thus, by
passing to a subsequence if necessary, the $x_i=(\tau_i, p)$
converge to some $x =(\tau, p)$. This means that, passing to a
further subsequence if necessary, the $\mu_i$ converge to a null
geodesic $\mu$ with future endpoint $x$, such that the projection
of $\mu$ onto $(\S_0, \wh)$ is the ray $\sigma$. Since the $q_i$
exhaust $\eta$ to the past, $\mu$ is past inextendible. If $u$
denotes an affine parameter along $\mu$ then it follows that $du =
KN^2d{\tilde s}$, $K= {\rm const.}$, where $d{\tilde s}$ is arc
length along $\sigma$.  Since $\sigma$ has infinite length, and
$N^2$ has a positive lower bound on $\S_0$, it follows that $\mu$
is past complete in $(\S_0, h, N)$.

Finally, we show that $\mu$ is in fact a generator of $S^+$ with
future endpoint $q$. Since $\mu_n \subset J^+(\eta)$, this means
that $\mu \subset \overline{J^+(\eta)}=\overline{I^+(\eta)}$. Let
us assume that $\mu \cap I^+(\eta) \neq \phi$. Thus, there exists
an $a \in \mu$ and $b \in \eta$ such that $a \in I^+(b)$. Consider
a neighbourhood $U_a$ of $a$ such that $U_a\subset I^+(b)$. Since
$a \in \mu$ and $\mu$ is a limit curve to the $\mu_i$ there exists
an $N_1$ such that $\forall i> N_1$, there exists an $a_i \in
\mu_i$ such that $a_i\in U_a$. Moreover, since the $q_i$ exhaust
$\eta$ to the past, for any $b \in \eta$, $\exists \, \, N_2$ such
that $\forall i>N_2$, $q_i$ is to the past of $b$ on $\eta$. Thus,
$\exists \,\,N$ such that $\forall i>N$, $q_i$ lies to the past of
$b$ on $\eta$ and $a_i \in I^+(b)$ which implies that $a_i \in
I^+(q_i)$, which is not possible since $\mu_i$ is achronal. Hence
$\mu \subset \pd J^+(\eta)$. That its future endpoint $x=q$
follows, since $ x \in S^+$ and $x$ and $q$ lie along the same
orbit of the time-like Killing vector field $(\frac{\pd}{\pd
t})^a$.  A similar argument shows that the null generators of
$S^-=\pd J^-(\eta)$ are future complete. \qed

\noindent {\bf Remark.} If, further, $R_{ab}X^aX^b\ge 0$ for all 
null vectors $X^a$ then the proof of the NST \cite{gjg} implies 
that $S^+=S^-=S$ is a smooth, edgeless, totally geodesic null
hypersurface.  This ultimately leads to a contradiction in the
presence of black holes as will be shown in the following theorem:

\noindent {\bf Theorem 1.3.} {\sl Let $M$ be an $n+1$ dimensional
($n\geq 3$) AL-AdS spacetime with static domain of outer
communications $(\Sigma, h,N)$ satisfying the null energy
condition. If $\pi_1(\Sigma)$ is infinite and if the asymptotic
condition (C) (Definition 2.1) is satisfied, then $(\Sigma, h, N)$
does not admit regular black holes, {\it i.e.}, $\pd \sh =\emptyset$.}

\noindent {\bf Proof.} Suppose to the contrary that
$(\Sigma, h, N)$ is conformally compactifiable with conformal
boundary $\pd\sinf$, and that it has a regular black hole horizon
$\pd\sh\ne \emptyset$, as described in Section 2.

Recall, by definition,  $\S'= \S\cup \pd \sinf\cup\pd\sh$ is a
compact Riemannian manifold-with-boundary.  We truncate $\S'$ 
by removing small collared neighbourhoods (foliated by constant lapse
slices)  $\cU_H\approx [0,\e]\times \sh$ and $\cU_{\8}\approx
[0,\e]\times \sinf$ of $\pd\sh$ and $\pd\sinf$, respectively. In
this way we obtain an open set $\S_0$ in $\S$ whose closure
$\overline\S_0\subset\S$ is a compact manifold with boundary
$\pd\overline\S_0 =\cV_A \cup\cV_B$, where $\cV_A$ is a constant
lapse slice near $\pd\sh$ and $\cV_B$ is a constant lapse slice
near $\pd\sinf$. By Lemma 2.2, we may assume that $\cV_A$ is
strictly convex.  By Condition C we may assume that $\cV_B$, and
all constant lapse slices near $\cV_B$, are weakly convex.

Let $(\overline \S_0^*, \wh^*)$ be the Riemannian universal cover
of $(\overline \S_0, \wh)$. Since $\overline \S_0$ is a
deformation retract of $\S'$, its fundamental group is isomorphic
to that of $\S'$, and hence to that of $\S$, and so $\vert\pi_1
(\overline \S_0)\vert = \8$. Thus, $(\overline \S_0^*, \wh^*)$ is a 
complete noncompact Riemannian manifold with boundary  $\pd\overline 
\S_0^* = \cV_A^*\cup\cV_B^*$, where $\cV_A^*$ covers $\cV_A$ and
$\cV_B^*$ covers $\cV_B$.  The convexity conditions on $\cV_A$ and
$\cV_B$ lift to $\cV_A^*$ and $\cV_B^*$.

$\S_0^*$, the manifold interior of $\overline \S_0^*$, corresponds
to the universal cover of $\S_0$. We now show that $(\S_0^*,\wh^*)$
contains a Fermat line. This construction is a simple extension of
that used in the Lorentzian structure theorem of \cite{gsw2}. Let
$p \in \S_0^*$ and $\{q_i\}\subset \S_0^*$ be a sequence of points
uniformly bounded away from $\pd\overline\S_0^*$ such that the
distance from $p$ to $q_i$ tends to infinity. For each $i$, let
$\gamma_i$ be a length minimizing geodesic in $\S_0^*$ from $p$ to
$q_i$ .  Our convexity conditions imply  that such  $\gamma_i$
exist, and are uniformly bounded away from $\pd\overline\S_0^*$.
Since $\overline\S_0$ is compact, the universal cover
$\overline\S_0^*$ admits a compact fundamental domain $D$. For
every point in  $\overline\S_0^*$ there exists a covering space
transformation, $\fg\in \pi_1(\overline\S_0^*)$, mapping it to a
point in $D$.  Let $\fg_i \in\pi_1(\overline\S_0^*) $ map the
midpoint $r_i$ of $\gamma_i$ into $D$.
Since the $\fg_i$ are isometries, the new curves
$\gamma_i'=\gamma_i\circ \fg_i$ are minimal geodesics which meet
$D$, and are bounded away from $\pd \overline\S_0^*$, since the
$\gamma_i$ are. Moreover, the lengths of the $\g_i'$ are now
unbounded in both directions. It follows that some subsequence of
$\{\g_i'\}$ converges to a complete line $\g$ in $(\S_0^*,\wh^*)$.

Consider the static spacetime $M_0^* = (\S_0^*, h^*,N^*)$, where
$N^* = N\circ\rho$, and\,\,\, $\rho : \S_0^* \to \S_0$ is the
covering map;  $M_0^*$ is the universal covering spacetime of $M_0
= (\S_0, h, N)$.  Now, $\g$ lifts to a complete null line $\eta$
in $(\S_0^*, h^*,N^*)$. By Lemma 3.2 and the remark following
its proof, the NST implies that $\eta$ is contained in  a smooth
totally geodesic, edgeless achronal null hypersurface $S$ in
$(\S_0^*,  h^*,N^*)$.  As $S$ is edgeless, and hence closed as a
subset of $(\S_0^*, h^*,N^*)$, and as its generators are
complete, it can be shown that $S$ maps diffeomorphically onto
$\S_0^*$ (viewed as the slice $t=0$) via the flow lines of 
the timelike Killing field
$(\frac{\pd}{\pd t})^a$; cf., \cite{gsw2} for details.  The null
generators of $S$, being  complete null lines, project to complete
Fermat lines in $\S_0^*$.  It follows that $(\S_0^*,\wh^*)$ is
ruled by complete Fermat lines.  Fix a point $x$ on the boundary
component $\cV_A$, and let $\{x_i\}$ be a sequence of points in
$\S_0^*$ such that $x_i\to x$.  Let $\s_i$ be a complete Fermat
line in $\S_0^*$ passing through $x_i$.  By passing to a
subsequence if necessary, $\s_i$ will converge to a complete
Fermat geodesic $\s$ contained in $\overline\S_0^*$ passing
tangentially through $x\in \cV_A$.  But this contradicts the fact
that $\cV_A$ is strictly convex. Thus we conclude that $(\Sigma,
h, N)$ does not admit regular black holes. \qed

\section{Applications: Proofs of Propositions 1.1 and 1.2}
\setcounter{equation}{0}

In special cases, condition (C) can be weakened to condition (S)
by using a boundary analysis along the lines of \cite{FG,gsw2}
where one considers a  foliation of  a neighbourhood of scri with
constant lapse surfaces. Consider a collar neighbourhood $\cO$ of
the boundary $\pd\sinf$ at $x\equiv\tN=0$. In the coordinates
$x^1=x,x^2,  \cdots, x^n$ the Fermat metric $\wh$ takes the form
\begin{equation}
\label{A4} \wh = \tW^{-1} dx^2 + \tb_{\a\b}dx^\a dx^\b,
\end{equation}
where $\tb_{\a\b}=\tb_{\a \b}(x,x^\gamma)$ is the induced metric
on the constant $x$ slices $\cV_x$, which are diffeomorphic to
$\pd \sinf$.

Let us first consider the class of spacetimes with Ricci flat
scri, $\pd \sinf=\re^{n-1}/\Gamma$, with $\Gamma$ a discrete
co-compact isometry group of $\re^{n-1}$. The boundary analysis of
\cite{gsw2} shows that the mean curvature of a constant lapse
surface in a neighbourhood of scri is
\begin{equation}
\tH(x)=-n x^{n-1} \mu +{\cal O}(x^n),  \label{C1}
\end{equation}
where the mass aspect $\mu$ is related to the the Ashtekar-Magnon
mass \cite{am} via
\begin{equation}
M_{AM}=\frac{1}{16\pi } \int_{\partial \tSigma} \mu \sqrt{\tb}\
dS,\label{C2}
\end{equation}
When the mass aspect is pointwise negative, the mean curvature is
positive. This, along with Condition (S) implies Condition (C).
Then it is an immediate corollary of Theorem 3.1 that:

\noindent {\bf Proposition 1.1.} {\sl Let $M$  be an $n+1$
dimensional ($n\geq 3$)  AL-AdS spacetime, with static domain of
outer communications $(\Sigma, h, N)$, which satisfies the null
energy condition and has a Ricci flat scri, and whose
Ashtekar-Magnon mass aspect on scri is pointwise negative.  If
$\pi_1(\Sigma)$ is infinite and if the asymptotic condition (S) is
satisfied, then $(\Sigma, h, N)$ does not admit regular 
black holes, {\it i.e.}, $\pd \sh =\emptyset$.}

Next, we consider the case in which $\pd\sinf$ has positive scalar
curvature in the induced metric $\tb_{ab}(0)$. In the collar
neighbourhood $\cO$ of $\pd\sinf$, the mean curvature of the
$\cV_x$ can be expressed as
\begin{equation}\label{A17}
\tH(x)= x\pd_x \tH(0) + O(x^2),
\end{equation}
where we have used the fact that the extrinsic curvature
$\tH_{\a\b}(0)=0$ (see (\ref{A6t})).  By taking the projection of
the field equation (\ref{A1}) tangent to $\cV_x$ and
differentiating once (i.e. putting $k=1$ in (\ref{A7})) we obtain
\begin{equation} \label{A15}
\pd_x \tH_{\a\b}(0) = \frac{\ell}{(n-2)}\cR_{\a\b}(0),
\end{equation}
where $\cR_{\a\b}$ is the Ricci curvature with respect to
$\tb_{\a\b}$. Tracing the above,  and substituting into
(\ref{A17}) we obtain
\begin{equation}
\tH(x)= \frac{\ell}{(n-2)}\, \tS \, \, x + O(x^2),
\end{equation}
where $\tS$ is the scalar curvature $\pd\sinf$ with respect to
$\tb_{\a\b}$. $\tH(x)$  is clearly positive for small enough $x$,
under the assumption $\tS>0$. Thus, Condition (C) reduces to
Condition (S) in this case, and we obtain another corollary of Theorem 
1.3:

\noindent {\bf Proposition 1.2.} {\sl Let $M$  be an $n+1$
dimensional ($n\geq 3$)  AL-AdS spacetime, with static domain of
outer communications $(\Sigma, h, N)$ satisfying the null energy
condition. Suppose (a) the conformal boundary at infinity $\pd
\sinf$ has positive scalar curvature with respect to the Fermat 
metric, (b) the fundamental group of $\S$ is infinite, 
$\vert \pi_1(\S)\vert = \8$, and (c) the asymptotic
condition (S) is satisfied. Then $(\Sigma, h, N)$ does 
not admit regular black holes.}

To conclude this section, we examine Condition (S) for a class of
conformal boundaries $\pd\sinf$ which naturally have positive
scalar curvature and infinite fundamental group, such as
$S^2\times S^1$.

Consider the class of spacetimes with ``mixed'' boundary $\pd
\sinf = U \times V$, such that $U$ and $V$ are compact, with ${\rm
dim}(U) \geq 2$ and ${\rm dim}(V)\geq 1$. In a collar
neighbourhood of $\pd\sinf$ the constant lapse surfaces $\cV_x$
are diffeomorphic to $U\times V$. Let $\u(\vec u, \vec v,x)$ and
$\v(\vec u, \vec v,x)$ be the induced metrics with respect to
$\tb_{\a\b}$ on $U$ and $V$ respectively, where $\vec u$ and $\vec
v$ are the coordinates on $U$ and $V$ respectively. We further
assume that  metric has the simple product form at $x=0$
\begin{equation}
\label{A9} \tb_{\a\b}(\vec u,\vec v, 0) = {\u}_{\a\b}(\vec u, 0) +
{\v}_{\a\b}(\vec v, 0),
\end{equation}
such that
\begin{equation}
{}^u\cR_{\a\b} (\vec u, 0)= \frac{n(n-1)}{2\ell^2} \u_{\a\b}(\vec
u, 0) ; \quad {}^\v\cR_{\a\b}(\vec v, 0)=0. \label{C3}
\end{equation}
where ${}^\u\cR_{\a\b}$ and ${}^\v\cR_{\a\b}$ are the Ricci
tensors associated with $\u$ and $\v$, respectively. Our boundary
data are thus ``mixed'' in the sense that it is a product of a
positive Ricci curvature Einstein manifold and a zero curvature
Einstein manifold. We leave the details of the boundary analysis
for this class of boundary data to the appendix, only stating the
relevant results here.

{}From (\ref{A15}) and (\ref{C3}), the projection along $U$ of the 
second fundamental form $\tP_{\delta \rho}=\u^\a_{\,\,\,\delta}
\u^\b_{\,\,\, \rho}\tH_{\a \b}$ of $\cV_x$ has the expansion
\begin{equation}
\tP_{\a \b}(\vec u,\vec v, x)=\frac{n(n-1)}{2 (n-2) \, \ell}\, \,
x \, \, \u_{\a \b}(\vec u, 0) + O(x^2),
\end{equation}
which, by itself, for small enough $x$, has positive eigenvalues.
{}From Proposition A.3, the remaining
components of the second fundamental form $\tQ_{\delta
\beta}=\v^\a_{\,\,\, \delta}\tH_{\a \b}$ can be expanded as
\begin{equation}
\tQ_{\a\b}(\vec u,\vec v, x) = \frac{1}{(n-1)!}\, \,
x^{n-1}\tQ^{(n-1)}_{\a\b}(\vec u, \vec v, 0) + O(x^n).
\end{equation}
Since the boundary data $\tH^{(n-1)}_{\a\b}(\vec u, \vec v, 0)$ are
independent of $\tb_{\a\b}(\vec u, \vec v, 0)$, we are free to choose
$\tQ^{(n-1)}_{\a\b}(\vec u, \vec v, 0)$ so that $\tH_{ab}$ has only
non-negative eigenvalues in a small enough neighbourhood of scri.
Thus, Condition (S) can indeed be satisfied in a neighbourhood of the
conformal boundary when it is of the type under consideration here.
\footnote{If the ``mixed'' boundary were
instead such that ${}^\v\cR_{\a\b}(\vec v, 0)$ was Einstein with
{\it negative} curvature, then it can be seen that Condition (S)
could never be satisfied.}

\section*{Acknowledgements}
\setcounter{equation}{0}

\noindent This work was partially supported by grants from the
National Science Foundation (USA) DMS-0104042 and the Natural
Sciences and Engineering Research Council (Canada). SS was
partially supported by a postdoctoral fellowship from the Pacific
Institute for the Mathematical Sciences.

\appendix
\setcounter{equation}{0}

\section{Boundary analysis}

Before we proceed, it is useful to compare our analysis with that
of \cite{FG}. While the work of \cite{FG} is general and includes
metrics of both Lorentzian and Euclidean signature, it needs to be
modified to suit the purpose of our work.  In \cite{FG} a
particular choice of conformal gauge is made, namely,
$h=y^{-2}\wh_1={y^{-2}}(dy^2 \oplus \tb_1)$ and the boundary data
$\{ \pd_y^k \tb_1(0)\} $ is then analysed.  However, the choice of
a Fermat conformal gauge is crucial to the method employed in our
work, since we specifically require conditions (C) and (S)  to be
satisfied with respect to the Fermat metric.  From (\ref{A4}) it
is clear that the Fermat conformal gauge coincides with the
conformal gauge used in \cite{FG} only when $\tW =1$. Moreover,
while  $\tb(0)=\tb_1(0)$ the boundary data $\{\pd_x^k\tb(0)\} $
and $\{\pd_y^k\tb_1(0)\}$ will in general differ.

{}From Definition 2.1 and (\ref{A4}) $x^{1-n} T_{ab}$ should have a
 smooth limit to scri. Thus in a neighbourhood of scri we have
\begin{equation} \label{A5}
T_{ab}(x,x^\alpha) = A_{ab}(x^\alpha)x^{n-1} +O(x^n).
\end{equation}
Moreover, the  second fundamental form of $\cV_x$ can be expressed
as
\begin{equation}
\label{A6} \tH_{\a \b} = -\frac{1}{2\psi}\pd_x \tb_{\a \b},
\end{equation}
where $\psi=\frac{1}{\sqrt{\tW}}$.  Taking the projections of
(\ref{A1}) tangent and normal to the $\cV_x$ we obtain
\begin{eqnarray}
\tR_{\a\b}&=& \frac{(n-1)}{x\psi} \tH_{\a\b} + 8 \pi
[\cA_{\a\b}x^{n-1} + \cO(x^n)],  \\
\tR_{xx}&=& \frac{(n-1)}{x\psi} \pd_x \psi + 8 \pi [\cA_{xx}
x^{n-1} + \cO(x^n)],
\end{eqnarray}
where we have used (\ref{A5}) and defined $\cA_{ab}= A_{ab} -
\frac{1}{n-1}\wh_{ab}\wh^{cd}A_{cd}$.  As in \cite{gsw2}  we may
expand the left-hand sides of the above equations to obtain
\begin{eqnarray}
\tH_{\a\b}&=&\frac{x}{n-1}\biggl(\pd_x \tH_{\a\b} + 2 \tH_{\a
\gamma}\tH_\b^\gamma \psi - \tH \tH_{\a \b}\psi  + \cR_{\a\b} \psi
- D_\a D_\b \psi \nonumber \\ && + 8 \pi \psi [\cA_{\a\b}x^{n-1} +
\cO(x^n)]\biggr), \label{A4t} \\
\pd_x \psi &=& \frac{x}{n-1} \psi^2\biggl(\pd_x \tH - D^2 \psi -
\tH_{\a\b} \tH^{\a\b}\psi  + \frac{1}{\psi^2} 8 \pi
[\cA_{xx}x^{n-1} + \cO(x^n)]\biggr), \nonumber \\ \label{A5t}
\end{eqnarray}
where $D_\a$ is the connection compatible with $\tb_{\a\b}$, and
$\cR_{\a\b}$ is the associated Ricci curvature. Assuming the
conformal metric is  $C^2$ at $x=0$, this yields
\begin{equation}
\label{A6t} \tH_{\a\b}(0) = \pd_x \tb_{\a \b}(0)= \pd_x \psi(0) =
0; \quad \psi(0)=\ell
\end{equation}

Assuming  $C^k$ regularity of the metric for $ 1\leq k < n-1$,
(\ref{A4t}) and (\ref{A5t}) may be differentiated $k$ times to
obtain
\begin{eqnarray}
\tH_{\a \b}^{(k)}(0)&=& \frac{k}{(n-k-1)}\pd_x^{k-1} (2
\tH_{\a\gamma} \tH_{\b}^\gamma \psi - \tH \tH_{\a\b}\psi + \psi
\cR_{\a\b} - D_\a D_\b \psi)\bigg \vert_0 \nonumber \\
\label{A7}\\ \psi^{(k+1)}(0) &=& \frac{k}{n-1} \pd_x^{k-1}( \psi^2
\pd_x \tH -\psi^2 D^2 \psi -\psi^3 \tH_{\a \b} \tH^{\a \b})\bigg
\vert_0, \nonumber \\ \label{A8}
\end{eqnarray}
where $X^{(k)}\equiv \pd_x^k X \equiv\pd^k X/\pd x^k$ and
$X^{(0)}=X$. Notice that the matter stress-energy terms do not
contribute to these expressions because of the assumed fall-off
rates (\ref{A5}).

It is handy to write down the expressions relating
$\tH^{(k)}_{\a\b}(0)$ and $\tb^{(k+1)}_{\a\b}(0)$, which we obtain
by differentiating (\ref{A6}) $k$ times and using (\ref{A6t}),
\begin{equation}
\label{A10} \tH_{\a\b}^{(k)}(0)=-\frac{1}{2} \sum_{i=0}^{k-1}
\binom{k}{i} \, \, \, \pd_x^i(\psi^{-1})(0)
\tb_{\a\b}^{(k+1-i)}(0),
\end{equation}
or,
\begin{equation}\label{A11}
\tb_{\a\b}^{(k+1)}(0)=-2\sum_{i=0}^{k-1} \binom{k}{i}\,\,\,
\psi^{(i)}(0) \tH_{\a\b}^{(k-i)}(0).
\end{equation}
By inspection,  we notice that

\noindent {\bf Lemma A.1.} {\sl For $k>0$ even, every term in
(\ref{A10}) will contain either a $\tb^{(m+2)}_{\a\b}(0)$, or a
$\psi^{(m)}(0)$ where $m$ is odd and $1\leq m \leq k-1$.
Similarly, for $k$ even, every term in (\ref{A11}) will contain
either  an $\tH^{(m+1)}_{\a\b}(0)$, or a $\psi^{(m)}(0)$ where $m$
is odd and $1\leq m \leq k-1$.}

\noindent {\bf Proposition A.2} {\sl Assume that $(\Sigma, h, N)$
is a $C^{n}$ differentiable static spacetime which is AL-AdS. For
even $k < n-1$,
$\tb^{(k+1)}_{\a\b}(0)=\tH^{(k)}_{\a\b}(0)=\psi^{(k+1)}(0)=0$. For
$k=n-1$, $\tb^{(k+1)}_{\a\b}(0)$ and $\tH^{(k)}_{\a\b}(0)$ cannot
be determined from $\tb_{\a\b}(0)$ and constitute independent
boundary data.}

\noindent {\bf Remark.} This matches the results of \cite{FG} for
their choice of conformal gauge. Notice that unlike \cite{gsw2} we
do not require that the conformal boundary be Ricci flat.

\noindent {\bf Proof.} As in \cite{FG,gsw2} the proof is
iterative. We take $k$ to be even and $< n-1$ throughout the
proof.

\noindent {\bf Step A:} Consider the expression for
$\psi^{(k+1)}(0)$ in (\ref{A8}). By inspection, and using Lemma
A.1 and the Leibniz rule, every term in the expression can be seen
to contain either a (i) $\psi^{(m)}(0)$ for $m$ odd and $1 \leq m
\leq k-1$ or a (ii) $\tb^{(l)}_{\a\b}(0)$ for $l$ odd and $1\leq
l\leq k+1$ (or both). Applying this recursively to
$\psi^{(m)}(0)$, and using $\pd_x \psi (0)=0$ from (\ref{A6t}),
this simplifies to the fact that every term in the right hand side
of (\ref{A8}) must contain a $\tb_{\a\b}^{(m)}(0)$ for $m$ odd and
$1\leq m\leq k+1$.

\noindent {\bf Step B:} Next, consider the expression for
$\tH^{(k)}_{\a\b}(0)$ in (\ref{A7}). Again, using the Leibniz rule
and Lemma A.1, we see that every term in the expression contains
at least a (i) $\psi^{(m)}(0)$ or a (ii)
$\tb^{(m)}_{\delta\rho}(0)$ for $m$ odd and $m\leq k-1$. From step
A above, this implies that every term in the expansion for
$\tH^{(k)}_{\a\b}$ contains a $\tb_{\delta\rho}^{(m)}(0)$ for $m$
odd and $1\leq m\leq k-1$. Using this recursively with Lemma A.1,
this means that $\tb^{(k+1)}_{\a\b}$ can be expanded such that
each term in the expansion contains a $\tb_{\delta\rho}^{(m)}(0)$
for $m$ odd and $1\leq m\leq k-1$.

Thus, if $\tb_{\delta\rho}^{(m)}(0)=0$ $\forall \, \, m$ odd and
$1\leq m\leq k-1$, then from steps A and B we may conclude that
$\tb_{\a\b}^{(k+1)}(0)=0$ $\forall \,\, k$ even and $0<k<n-1$.
Since this is true for $k=0$, i.e., $\tb^{(1)}_{\a\b}=0$
(\ref{A6t}), by iteration we may conclude that
$\tb_{\gamma\delta}^{(k+1)}(0)=0$ $\forall \,\, k$ even and
$0<k<n-1$. It follows from step A, and Lemma A.1 that
$\psi^{(k+1)}(0)=\tH^{(k)}_{\a\b}(0)=0$ $\forall \,\, k$ even and
$0<k<n-1$. \qed.

Next, we consider the class of boundaries (\ref{A9}, \ref{C3}).
However, to prove  Proposition A.3 below, it suffices to merely
require that
\begin{equation}
{}^\v\cR_{\a\b}(0)=0, \label{C3g}
\end{equation}
with $\u(\vec u, 0)$ left arbitrary. As in Section 4, we will
use the shorthand
\begin{equation}
\tP_{\a\b}(\vec u, \vec v, x)= (\u^{\delta}_{\,\,\, \a}
\u^\rho_{\,\,\, \b}\tH_{\delta \rho})(\vec u, \vec v, x)
\end{equation}
to denote the projection of the second fundamental form of the
constant lapse surfaces along $U$, and
\begin{equation}
\tQ_{\a\b}(\vec u,\vec v, x) = (\v^\delta_{\,\,\, \a} \, \tH_{\b
\delta})(\vec u, \vec v, x).
\end{equation}
to denote the remaining components.

\noindent {\bf Proposition A.3.} {\sl Let $(\Sigma, h, N)$ be a
$C^{n}$ differentiable, static spacetime which is AL-AdS.  Let the
conformal boundary at $x=0$ be of the form described in (\ref{A9},
\ref{C3g}). Then, $\tQ_{\delta \b}^{(k)}(0)=0 $, $\forall \, \, k
< n-1$.}

\noindent {\bf Proof.} From Proposition A.2 we know that
$\tb^{(k+1)}_{\a\b}=\tH^{(k)}_{\a\b}(0)= 0$, $\forall k$ even and
$0 \leq k\leq n-1$. Thus, we need to only consider the case $k$
odd. For the rest of the proof we will assume that $k$ is odd and
$k<n-1$.

\noindent {\bf Step A:} We begin by differentiating
$\tP^{(k)}_{\a\b}(0)$ along $V$, $\v\cdot (\pd \,
\tP^{(k)}_{\a\b})(0)$,
where $\v\cdot\pd\equiv \v^\a_{\,\,\,\gamma}\pd_\a$. Using the
Leibniz rule, Proposition A.2, (\ref{A10}), (\ref{A11}),
(\ref{A6t}) and (\ref{A9}), we see that every term on the right
hand side of the expression contains either a (i)
$\tQ^{(m-1)}_{\delta \rho}(0)$ or a (ii) $\v\cdot\pd \,
(\psi^{(m)})(0)$ or a (iii) $\v\cdot \pd \, (P^{(m)}_{\delta
\rho})(0)$ for $1\leq m \leq k-1$. Applying this recursively to
(iii), this implies that every term in the expression for
$\v\cdot\pd \,(\tP_{\a\b}^{(k)})(0)$ contains either a (i)
$\tQ^{(m-1)}_{\delta \rho}(0)$ or a (ii)$\v\cdot\pd \,
(\psi^{(m)})(0)$ for $1\leq m \leq k-1$.

\noindent {\bf Step B:} Next, differentiate (\ref{A8}) along $V$,
to obtain an expression for $\v\cdot\pd \, (\psi^{(k+1)})(0)$.
Again, using the Leibniz rule, Proposition A.2, (\ref{A10}),
(\ref{A11}), (\ref{A6t}) and (\ref{A9}), we see that every term in
the expression contains either a (i) $\tQ^{(m-1)}_{\delta \b}(0)$
or a (ii) $\v\cdot\pd\, (\tP^{(m)}_{\delta \eta}) (0)$ or a (iii)
$\v\cdot\pd (\psi^{(m)})(0)$ for $1 \leq m \leq k-1$. Applying
this recursively along with the results of step A, we see that
every term in the expression for $\v\cdot\pd \, (\psi^{(k+1)})(0)$
contains a $\tQ^{(m)}_{\delta \b}(0)$ for $1 \leq m \leq k-2$.

\noindent {\bf Step C:} Finally, we consider the projection of
(\ref{A7}) along $U$, $\tQ^{(k)}_{\delta \b}(0)$. Again, using the
Leibniz rule, Proposition A.2, (\ref{A10}), (\ref{A11}),
(\ref{A6t}) and (\ref{A9}), we see that every term in the
projection of the right hand side of (\ref{A7}) contains either a
(i)
 $\tQ^{(m-1)}_{\delta \b}(0)$
or a (ii)
 $\v\cdot \pd \, (P^{(m)}_{\delta \rho})(0)$
or a (iii) $\v\cdot\pd (\psi^{(m)})(0)$ for $1 \leq m \leq k-1$.
Applying this recursively along with the results of steps A and B
above, we see that every term in the expansion of
$\tQ^{(k)}_{\delta \b}(0)$ contains a $\tQ^{(m)}_{\delta \b}(0)$
for $1 \leq m \leq k-2$.

Thus, if $\tQ^{(m)}_{\a \b}(0)=0$ $\forall m$ satisfying $1\leq m
\leq k-2$ then $\tQ^{(k)}_{\a \b}(0)=0$.
{}From (\ref{A7}) and (\ref{A6t}), $\pd_x
\tH_{\a\b}(0)=\frac{\ell}{(n-2)} \cR_{\a\b}(0)$. Since
${}^\v\cR_{\a\b}(0)=0$ and $\v^\delta_{\,\, \,\a} \cR_{\delta
\b}(0)=0$ from (\ref{A9}), $\tQ^{(1)}_{\a
\b}(0)=\frac{\ell}{(n-2)} \, \, \v^\delta_{\,\, \,\a} \cR_{\delta
\b}(0)=0$. Since  $\tQ^{(k)}_{\a\b}(0)=0$ for $k=1$, by induction,
we can conclude that $\tQ^{(k)}_{\a \b}(0)=0$  $\forall k < n-1$.
\qed.

\end{document}